# Epileptic Seizure Detection: A Deep Learning Approach

Ramy Hussein*§, Hamid Palangi†, Rabab Ward§, and Z. Jane Wang§

*Abstract*—Epilepsy is the second most common brain disorder after migraine. Automatic detection of epileptic seizures can considerably improve the patients' quality of life. Current Electroencephalogram (EEG)-based seizure detection systems encounter many challenges in real-life situations. The EEGs are non-stationary signals and seizure patterns vary across patients and recording sessions. Moreover, EEG data are prone to numerous noise types that negatively affect the detection accuracy of epileptic seizures. To address these challenges, we introduce the use of a deep learning-based approach that automatically learns the discriminative EEG features of epileptic seizures. Specifically, to reveal the correlation between successive data samples, the time-series EEG data are first segmented into a sequence of non-overlapping epochs. Second, Long Short-Term Memory (LSTM) network is used to learn the high-level representations of the normal and the seizure EEG patterns. Third, these representations are fed into Softmax function for training and classification. The results on a well-known benchmark clinical dataset demonstrate the superiority of the proposed approach over the existing state-of-the-art methods. Furthermore, our approach is shown to be robust in noisy and real-life conditions. Compared to current methods that are quite sensitive to noise, the proposed method maintains its high detection performance in the presence of common EEG artifacts (muscle activities and eye-blinking) as well as white noise.

## I. INTRODUCTION

EPILEPSY is a chronic neurological disorder of the brain that affects people of all ages. Approximately 70 million people worldwide have epilepsy, making it the second most common neurological diseases after migraine [1]. The defining characteristic of epilepsy is recurrent seizures that strike without warning. Symptoms may range from brief suspension of awareness to violent convulsions and sometimes loss of consciousness [2]. Epileptic seizure detection plays a key role in improving the quality of life of epileptic patients. Electroencephalogram (EEG) is the prime signal that has been widely used for the diagnosis of epilepsy. The visual inspection of EEG is unfortunately labour- and time-consuming. Also, around 75% of people with epilepsy live in low- and middle-income countries and cannot afford consulting neurologists or practitioners [3]. Those limitations have encouraged scholars to develop automatic EEG-based seizure detection systems.

A vast number of methods have been developed for automatic seizure detection using EEG signals. Extracting features that best describe the behaviour of EEGs is of great importance for automatic seizure detection systems' performance. Several feature extraction and selection techniques have been reported in the literature. Most of them use hand-wrought features in the time-domain [4], [5], frequency-domain [6]–[8], time-frequency domain [9]–[12] or sometimes in a combination of two domains [13]. However, these domain-based methods encounter three main challenges. First, they are sensitive (not robust enough) to acute variations in seizure patterns. This is because the EEG data is non-stationary and its statistical features change across different subjects and over time for the same subject. Secondly, EEG data acquisition systems are very susceptible to a diverse range of artifacts such as muscle activities, eye-blinks, and environmental white noise. All these sources of noise can alter the genuine EEG features and hence seriously affect the performance accuracy of seizure detection systems. The authors of [14] have studied the impact of high noise levels on the recognition performance of epileptic seizures. It is worth highlighting that detecting seizures from noisy EEG data corrupted with a medium-level noise has resulted in a drop of 10% in the seizure detection accuracy [14]. Finally, most existing seizure detection systems have been trained on small-scale EEG datasets collected from few specific patients, making them less practical in clinical applications.

To address these limitations, we introduce a robust deep learning approach for automatic detection of epileptic seizures. Because the start of a seizure pattern emerges at random in the EEG signals, we first divide the time-series EEGs into short-length segments. This pre-processing step captures the temporal correlations among successive EEG data samples. We then feed these EEG segments into a recurrent neural network with long short-term memory cells to learn the most robust and discriminative EEG features for epileptic seizure detection. The learned features are then fed into a softmax classifier layer which calculates the cross-entropy between true labels and predicted labels for the data. We apply the proposed model to the well-known benchmark dataset provided by Bonn University [15]. We first examine its detection performance under ideal conditions, i.e., when the EEG data are completely free of noise. Results show that our approach achieves superior detection performance relative to several state-of-the-art methods listed in Secion V. Moreover, the proposed model is inspected under real-life conditions, where

The first author is funded by Vanier Canada Graduate Scholarship from the Natural Sciences and Engineering Research Council of Canada (NSERC).
*Corresponding Author: ramy@ece.ubc.ca
§Ramy Hussein, Z. Jane Wang and Rabab Ward are with the Department of Electrical and Computer Engineering, University of British Columbia, Vancouver, BC V6T 1Z4, Canada.
†Hamid Palangi is with Microsoft Research AI, Redmond, WA 98052, United States.

the EEG data are corrupted with three different sources of noise: muscle artifacts, eyes movement, and environmental noise. Our approach is proven to be robust against all these types of artifacts. It maintains high detection accuracies at different noise levels, making it more relevant to clinical applications. Other state-of-the-art methods studied in this work, are not as robust to these artifacts and noise levels.

## II. DATASET

### A. Description of EEG Dataset.

In this study, we conduct our seizure detection experiments on the publicly available EEG dataset provided by Bonn University [15]. To the best of our knowledge, this is the most widely used dataset for epileptic seizure detection. It includes five different sets denoted A, B, C, D, and E; each includes 100 single-channel EEG signals of 23.6 seconds duration. Sets A and B contain surface EEG signals recorded from 5 healthy participants using the standardized 10-20 system for EEG electrode placement [16]. During the recording, participants were awake and relaxed with eyes open (Set A) and eyes closed (Set B). Sets C and D consist of intracranial EEG signals taken from five epileptic patients during seizure-free intervals. The EEG signals in set C are recorded using electrodes implanted in the brain epileptogenic zone, while those in set D are recorded from the hippocampal formation of the opposite hemisphere of the brain. Set E includes EEG segments recorded from five epileptic patients while experiencing active seizures.

All the EEG signals are sampled at 173.6Hz and digitized using a 12-bit analog-to-digital converter. The EEG data provided by the Bonn Dataset does not have artifacts. Prior to publishing the dataset, the captured EEG segments containing artifacts had been deleted and those containing delicate artifacts had been denoised using a band-pass filter with cut-off frequencies of 0.53Hz and 40Hz.

### B. Common EEG Artifacts.

In practice, EEG recordings are often corrupted with several types of artifacts. These artifacts may negatively affect the genuine manifestations of seizure patterns and severely influence the detection accuracy of epileptic seizures. The authors of [17] reviewed the most common types of EEG artifacts and developed models that mimic their behaviour. In this paper, we used these models to study the most three vital and inevitable sources of artifacts, which are:

1) Muscle Artifacts: As depicted in [17], muscle activities can be modeled by random noise filtered with a band-pass filter (BPF) of 20Hz and 60Hz cut-off frequencies and multiplied by a typical muscle scalp map.
2) Eyes Movement/Blinking: The eye blinks can be modeled as a random noise signal filtered with a BPF of 1Hz and 3Hz cut-off frequencies [17].
3) White Noise: The electrical and environmental noise are modeled as additive white Gaussian noise [17].

Figure 1(a) shows an arbitrary noise-free EEG signal from set A, while Figures 1(b), (c), and (d) show the corrupted versions of the same signal after adding muscle artifacts, eye-blinking, and white noise, respectively. Figures 1(e), (f), (g) and (h) also depict the frequency spectra of the time-series EEG signals shown in Figures 1(a), (b), (c) and (d), respectively. The amplitudes of the muscle artifacts, eye-blinking, and white noise can be adjusted to produce noisy EEG signals with different signal-to-noise-ratios (SNRs). The SNR of the noisy signals shown in Figure 1 is set to 0dB, this is where the noise signal have the same power as the EEG signal. Matlab$^{TM}$ software was used to generate the synthetic artifacts and add them to the clean EEG data.

## III. RELATED WORK

The problem of EEG-based epileptic seizure detection has been broadly investigated over the past three decades. The published work can be sorted into three main classification problems. The first problem is to differentiate between two distinct classes; *Normal (set A)* and *Ictal (set E)* EEG patterns [18]–[35]. The second problem is to differentiate between *Normal (set A)*, *Inter-ictal (set C)*, and *Ictal (set E)* EEG patterns [36]–[48]. The third and most challenging problem addresses the discrimination between the five different EEG sets; *A, B, C, D, and E* [49]–[54]. It is worth highlighting that none of the studies below in this section take into consideration the existence of artifacts and their negative influence on the seizure detection accuracy.

### A. Two-class EEG Classification.

Most of the two-class seizure detection problems focus on the classification between normal EEG segments taken from healthy persons (set A) and seizure EEG patterns taken from epileptic patients while experiencing active seizures (set E) [18]–[28]. Aarabi *et al.* proposed an automated seizure detection system using a set of representative EEG features extracted from time domain, frequency domain and wavelet domain as well as auto-regressive coefficients and cepstral features [18]. All these features were fed altogether into a back-propagation neural network (BNN) classifier with two hidden layers and resulted in an average classification accuracy of 93.00%. In [19], Subasi *et al.* used wavelet transform to derive the EEG frequency bands and then use all the spectral components as an input to the mixture of experts (ME) classifier; an average classification accuracy of 94.50% was achieved. Polat *et al.* achieved a higher classification accuracy of 98.68% using a decision tree (DT) classifier [20].

Furthermore, Chandaka *et al.* used the EEG cross-correlation coefficients to compute three statistical features, and hence present them as a feature vector to the support vector machine (SVM) for EEG classification [21]. This model yielded a modest seizure detection accuracy of 95.96%. Yuan *et al.* obtained comparable detection accuracies using the extreme learning machine (ELM) classifier and a set of non-linear features such as approximate entropy and Hurst exponent [22]. Wavelet transform was also used in [23] to analyze the EEG signals into five approximation and detail sub-bands. Then, the wavelet coefficients located in the low frequency range of 0-32Hz were used to compute the EEG features of energy and normalized coefficients. The linear discriminant analysis (LDA) classifier was used to prove the potential of the extracted features in detecting seizure onsets with a classification accuracy of 91.80%. In addition, the authors of [24] leveraged the permutation entropy as a delegate EEG feature for automatic detection of epileptic seizure. A SVM was utilized to differentiate between normal and epileptic EEG epochs; a 93.80% classification accuracy was achieved. Zhou *et al.* studied the capability of Bayesian LDA (BLDA) model to attain better results [25], where it was trained and tested on the EEG features of lacunarity and fluctuation index to achieve a classification accuracy of 96.67%.

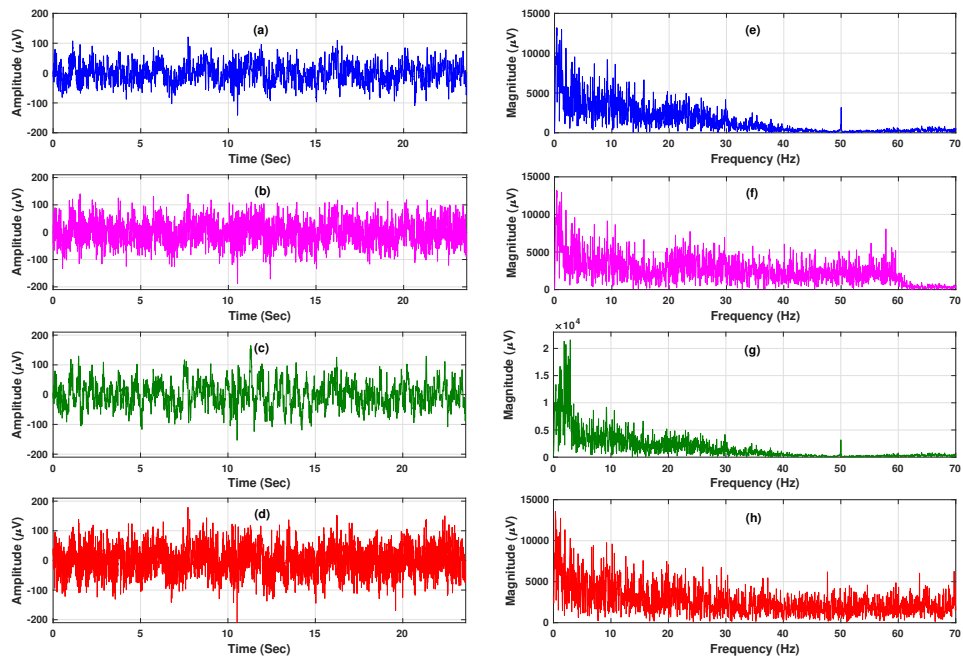

Figure 1. Time-series EEG signals and their corresponding spectra: (a) clean EEG example from set A; (b), (c), and (d) noisy EEG examples corrupted with muscle artifacts, eye-blinking, and white noise, respectively; (e-h) corresponding frequency spectra of (a-d), respectively.

Given the advantages of the wavelet transform outlined in the previous paragraph, it was also used in [26] to disband the EEG signals into five different frequency rhythms namely delta, theta, alpha, beta and gamma. A set of statistical and non-linear features was subsequently extracted from these rhythms and fed into a SVM classifier to achieve a superb detection accuracy of 97.50%. In [27], Song *et al.* also used the SVM together with the weighted permutation entropy features to obtain a classification accuracy of 97.25%. Furthermore, the multilevel wavelet transform was also used in [28] to decompose the EEG signals into a number of sub-bands, whose spectral features were extracted and used to construct the feature vector. As a consequence, the feature vector was introduced to the ELM for training and classification; promising results of 99.48% sensitivity was achieved.

A special case of the two-class problem is to differentiate between the seizure activities (set E) and any non-seizure activities (sets A, B, C or D). The main goal of this kind of problems is to accurately identify whether or not the patient experiences an active seizure. This can help patients, caregivers, and healthcare providers to administer the appropriate medication on time. In recent years, many researchers have shed the light on this particular problem [29]–[35], achieving high seizure detection accuracies. For instance, Guo *et al.* used the Wavelet-based approximate entropy features together with an artificial neural network (ANN) model to identify the seizure episodes with an average classification accuracy of 98.27% [29]. The authors of [30] developed a Genetic algorithm for automated EEG feature selection, that was used with k-nearest neighbors (KNN) classifier to boost the detection accuracy to 98.40%.

In 2013, the EEG signals were first analyzed using the approach of empirical mode decomposition (EMD) [31]. Four simple features were then extracted from the EEG decomposed components and fed into the KNN classifier for EEG classification; an average classification accuracy of 98.20% was achieved. In 2015, the authors of [32] used the same approach of EMD but with more robust features such as the spectral entropies and energies of EEG frequency bands. Using SVM, the classification accuracy was improved to 98.80%. In [33], Peker *et al.* used wavelet transform to analyze the EEG data into different rhythms and then computed five statistical features from each rhythm. These features are concatenated together and entered into the complex-valued neural networks (CVANN) classifier for seizure diagnosis. As a result, an average classification accuracy of 99.33% was achieved. Further, Jaiswal *et al.* presented a novel computationally-simple feature extraction technique named local neighbor descriptive pattern (LNDP) and they tested it along with different classification models including KNN, SVM, ANN and DT [34]. Experimental results show that the best detection performance can be fulfilled using LNDP jointly with the ANN classifier, where the highest classification accuracy of 98.72% is obtained. To further improve the seizure detection rate, a combination of time domain, frequency domain and time-frequency domain features were used together with SVM classifier to achieve the best classification rate of 99.25% [35].

*B. Three-class EEG Classification.*

This category of seizure detection problems addresses the classification of three different EEG classes: *Normal* EEG recorded from healthy volunteers, *Inter-ictal* EEG recorded from epileptic patients during seizure-free intervals and *Ictal* EEG recorded from epileptic patients while experiencing active seizures. Numerous relevant methods have been presented in the literature [36]–[48]. For example, the authors of [36] investigated the use of the recurrent neural network (RNN) as a classification model for epilepsy diagnosis. A satisfactory performance of 96.79% classification accuracy was achieved. In [37], Tzallas *et al.* reached a superior detection accuracy of 97.94% by using the ANN classifier together with the energy features of EEG frequency bands. Moreover, the work in [38] intoduced a novel classifier named radial basis function neural network (RBFNN), which was integrated with the wavelet features to achieve a seizure diagnostic accuracy of 96.60%. Furthermore, Übeyli *et al.* adopted wavelet transform to obtain and analyze the main spectral rhythms of the EEG signals [39]. Then, the statistical features that characterize the behavior of the EEGs were extracted and tested using the multilayer perceptron neural network (MLPNN) classifier. The results showed sensitivity, specificity, and classification accuracy of 96.00%, 94.00%, and 94.83%, respectively. In [40], a feature extraction method based on the sample entropy was used together with the ELM classifier and resulted in sensitivity, specificity, and classification accuracy of 97.26%, 98.77%, and 95.67%, respectively. Also, a set of temporal and spectral EEG features forming a more representative feature vector was fed into a MLPNN for EEG classification [41]. The epilepsy detection rates produced by this method were 97.46% for sensitivity, 98.74% for specificity, and 97.50% for classification accuracy.

In an effort to alleviate the computational complexity burden in seizure detection systems, Acharya *et al.* relaxed the need of any pre-processing techniques and worked directly on the raw EEG data [42], [43]. In [42], a set of robust EEG features including approximate entropy, sample entropy and phase entropy was computed from the recorded EEG signals and then fed into fuzzy Sugeno classifier (FSC) for EEG classification. This approach notably boosted the classification accuracy to 98.10%. In addition, Acharya *et al.* proposed, for the first time, the use of wavelet packet transform (WPT) to analyze the EEG signals into eight approximation and detail wavelet bands [43]. The wavelet coefficients of these bands were then used to infer the distinctive eigenvalues and use them as an input to the Gaussian mixture model (GMM) classifier, which in turn achieved an outstanding classification accuracy of 99.00%. An analogous classification accuracy of 98.67% was achieved in [44] by using a feature extraction method based on recurrence quantification analysis integrated with a two-stage classifier named error-correction output code (ECOC).

This approach notably boosted the classification accuracy to 98.10%. Additionally, Acharya *et al.* proposed, for the first time, the use of wavelet packet transform (WPT) to analyze the EEG signals into eight approximation and detail wavelet bands [43]. The wavelet coefficients of these bands were then used to infer the distinctive eigenvalues and use them as an input to the Gaussian mixture model (GMM) classifier, which in turn achieved an outstanding classification accuracy of 99.00%. An analogous classification accuracy of 98.67% was achieved in [44] by using a feature extraction method based on recurrence quantification analysis integrated with a two-stage classifier named error-correction output code (ECOC).

Further, the authors of [45] built a piecewise quadratic (PQ) classifier for detecting epileptic EEG episodes. They integrated this classifier with a combination of temporal, spectral, and non-linear features and reached up to 98.70% classification accuracy. Besides, in [46], a feature extraction method based on the discrete short-time Fourier transform was adopted together with a MLPNN classifier to discriminate between normal and seizure EEG epochs. As a result, the highest detection accuracy of 99.10% was achieved. Also, the independent component analysis (ICA) method was employed to determine the discriminatory features pertinent to epileptic seizures [47]. The extracted features together with the SVM classifier were used to achieve a sensitivity, specificity, and classification accuracy of 96.00%, 94.00%, and 95.00%, respectively. In [48], a seizure detection scheme based on some statistical features and a least-square SVM (LSSVM) classifier showed an average classification accuracy of 97.19% with a short computation time of 0.065 seconds.

*C. Five-class EEG Classification.*

This section addresses the classification of a data sample when the labels are one of five classes (which are *A, B, C, D, and E*. This kind of classification problems is more complex and harder to solve than the two-class and three-class problems. The main reason is that it attempts to differentiate between similar pathological EEG patterns corresponding to the same data class (e.g., the classification between EEG sets C and D, which are both Inter-ictal EEGs). But since the EEG sets of C and D are recorded from different epileptogenic brain zones [15], their correct classification holds a great potential in localizing the seizure foci inside the brain; making it quite advantageous for such kinds of vital applications. Here, we highlight the most recent work that handles such kinds of problems [49]–[54].

In [49], Güler *et al.* proposed one of the most efficient multi-class EEG classification methods for epileptic seizure detection. They extracted the best representative characteristics from the EEG wavelet coefficients and Lyapunov exponents. The probabilistic neural network (PNN) was used afterwards for EEG classification, where it achieved a notable classification accuracy of 98.05%. Also, Übeyli *et al.* developed an eigenvector-based method for EEG feature extraction, which in turn achieved a 99.30% classification accuracy using SVM [50]. In [51], the same authors used simple statistical features instead and a high classification accuracy of 99.20% was maintained.

Furthermore, the EEG spectral rhythms of delta, theta, alpha, beta, and gamma were also used in [52] as delegate features for EEG classification. Using these features, the multiclass SVM (MSVM) classifier attained a classification accuracy of 96.00%. Likewise, in [53], SVM was used in cooperation with the adaptive feature extraction method of wavelet approximate entropy and they together achieved a promising classification accuracy of 99.97%. Recently, Siuly *et al.* obtained the best classification accuracy ever [54]. They designed a novel statistical feature extraction scheme and integrated it with a MSVM to classify EEG signals; an impressive 99.99% classification accuracy was obtained.

## IV. METHODOLOGY

Deep learning has been proven to achieve promising results in different research problems such as face recognition [55], image classification [56], information retrieval [57] and speech recognition [58]. In this study, we propose the use of deep recurrent neural networks, particularly the long short-term memory (LSTM) model [59], for epileptic seizure diagnosis.

### A. High Level Picture

Figure 2 depicts the whole process of the proposed seizure detection system. The time-series EEG signals are first divided into smaller non-overlapping segments. These segments are then fed into the LSTM networks which are used for learning the high-level representations of the EEG signals. Next, the output of LSTM layer $U$ is presented as an input to the time-distributed Dense layer $h$ to find the most robust EEG features pertinent to epileptic seizures. Finally, a softmax layer is used to create the label predictions [60]. The detailed pipeline of the proposed approach is described in the following subsections.

We use the LSTM architecture illustrated in Figure 3 for the proposed seizure detection method. This figure has three gates (input, forget, output), a block input, a single cell (the Constant Error Carousel), an output activation function, and peephole connections [61]. The output of the block is recurrently connected back to the block input and all of the gates.

Let $\mathbf{x}^t$ be the input vector at time $t$, $B$ be the number of LSTM units and $M$ the number of inputs (EEG segments). Then we get the following weights for an LSTM layer:

- Input weights: $\mathbf{W}_z, \mathbf{W}_i, \mathbf{W}_f, \mathbf{W}_o \in \mathbb{R}^{B \times M}$
- Recurrent weights: $\mathbf{R}_z, \mathbf{R}_i, \mathbf{R}_f, \mathbf{R}_o \in \mathbb{R}^{B \times B}$
- Peephole weights: $\mathbf{P}_i, \mathbf{P}_f, \mathbf{P}_o \in \mathbb{R}^B$
- Bias weights: $\mathbf{b}_z, \mathbf{b}_i, \mathbf{b}_f, \mathbf{b}_o \in \mathbb{R}^B$

Considering Figure 3, the definitions of the vector relationships formulas for a basic LSTM layer forward pass can be written as [61]:

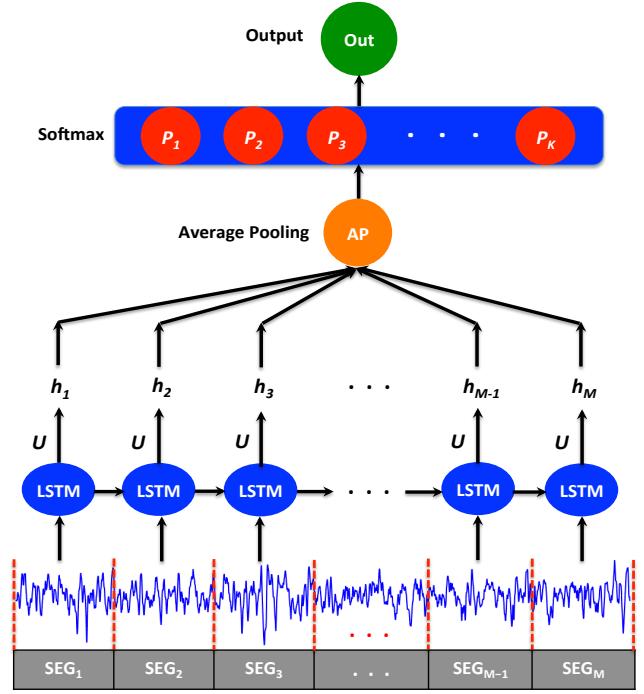

Figure 2. Schematic diagram of the overall seizure detection approach: $SEG_1$, $SEG_2$, $SEG_3$, $\cdots$, $SEG_{M-1}$, $SEG_M$ are corresponding to $1^{st}$, $2^{nd}$, $3^{rd}$, $\cdots$, $(M-1)^{th}$, and $M^{th}$ EEG segments of each EEG channel signal; LSTM stands for Long-Short-Term Memory; $U$ is the output of LSTM layer; $h_1$, $h_2$, $\cdots$, $h_M$ represent the Dense layer units; **AP** stands for the average pooling; $P_1$, $P_2$, $P_3$, $\cdots$, $P_K$ are the probabilities produced by softmax for the K-classes; **Out** stands for the output of the softmax layer (predicted label).

$$\bar{\mathbf{z}}^t = \mathbf{W}_z \mathbf{x}^t + \mathbf{R}_z \mathbf{y}^{t-1} + \mathbf{b}_z \tag{1}$$
$$\mathbf{z}^t = g(\bar{\mathbf{z}}^t) \qquad \text{block input} \tag{2}$$
$$\bar{\mathbf{i}}^t = \mathbf{W}_i \mathbf{x}^t + \mathbf{R}_i \mathbf{y}^{t-1} + \mathbf{P}_i \odot \mathbf{c}^{t-1} + \mathbf{b}_i \tag{3}$$
$$\mathbf{i}^t = \sigma(\bar{\mathbf{i}}^t) \qquad \text{input gate} \tag{4}$$
$$\bar{\mathbf{f}}^t = \mathbf{W}_f \mathbf{x}^t + \mathbf{R}_f \mathbf{y}^{t-1} + \mathbf{P}_f \odot \mathbf{c}^{t-1} + \mathbf{b}_f \tag{5}$$
$$\mathbf{f}^t = \sigma(\bar{\mathbf{f}}^t) \qquad \text{forget gate} \tag{6}$$
$$\mathbf{c}^t = \mathbf{z}^t \odot \mathbf{i}^t + \mathbf{c}^{t-1} \odot \mathbf{f}^t \qquad \text{cell} \tag{7}$$
$$\bar{\mathbf{o}}^t = \mathbf{W}_o \mathbf{x}^t + \mathbf{R}_o \mathbf{y}^{t-1} + \mathbf{P}_o \odot \mathbf{c}^t + \mathbf{b}_o \tag{8}$$
$$\mathbf{o}^t = \sigma(\bar{\mathbf{o}}^t) \qquad \text{output gate} \tag{9}$$
$$\mathbf{u}^t = h(\mathbf{c}^t) \odot \mathbf{o}^t \qquad \text{block output} \tag{10}$$

where $\sigma$, $g$, and $h$ are point-wise activation functions. The logistic sigmoid $\sigma(.)$ is used as a gate activation function and the hyperbolic tangent $g(.) = h(.) = tanh(.)$ is used as the input and output activation function of an LSTM unit. $\odot$ denotes the point-wise multiplication of two vectors [61].

### B. Proposed Method

#### 1) EEG Segmentation:

Biomedical data such as EEGs are usually non-stationary signals, *i.e.*, their statistical characteristics change over time [62]. The purpose of EEG segmentation is to divide a signal
4

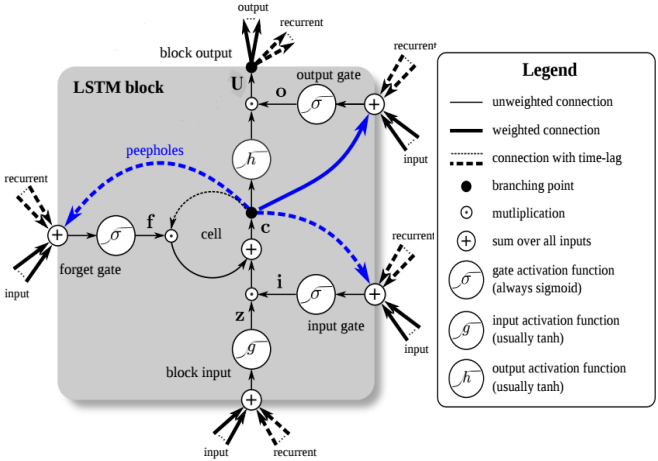

Figure 3. Detailed schematic of a Long-Short-Term Memory block [61].

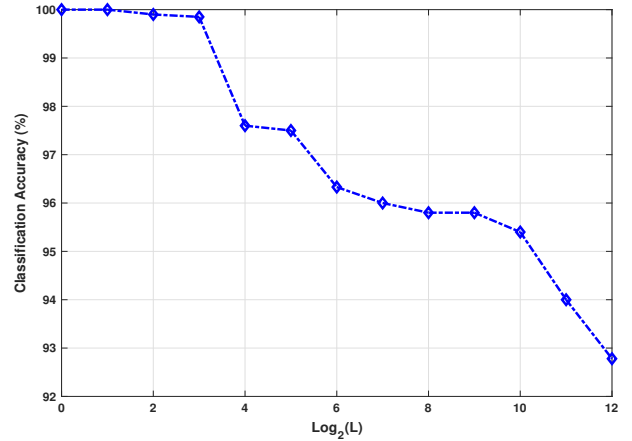

Figure 4. Classification accuracy against EEG segments' length.

into several pseudo-stationary epochs (segments) as these are expected to have similar statistical temporal and spectral features [63]. This is because the analysis of stationary signals is easier than non-stationary signals. Thus, EEG segmentation is usually applied as a pre-processing step for non-stationary signal analysis.

The other important factor behind EEG segmentation, particularly in this study, is the need to having a large number of labeled data samples. In general, it is hard to obtain sufficient well-labeled data for training deep neural networks in real life applications. The data segmentation, however, can help obtain more training samples, and hence improve the performance of the deep learning architecture under study. Over and above, EEG segmentation helps in finding the dependencies between the consecutive EEG data-points in each EEEG channel signal.

The EEG dataset under study includes 500 EEG signals, each of 23.6 seconds duration. And given the sampling rate of 173.6 Hz, the total number of data-points in each EEG signal, denoted by N, equals to 4096. All the EEG signals are devided into non-overlapping segments of a specific length (L). The most natural selection for L is L=1, i.e., having a predictive model like LSTM predicting sample 2 from sample 1, sample 3 from sample 2, and so on. This will be computationally slow in our study. To reduce computational complexity for a generic EEG segment length L, we create vectors of size L×1 and do all multiplications and additions in parallel for those L data-point vectors.

In our experiments, we tested a wide range of the EEG segment length and we inferred that increasing this length can lessen the computational cost of the LSTM models, but at the cost of detection accuracy [64]. Figure 4 depicts how the seizure detection accuracy decays with longer segment lengths. It also shows that L=1 and L=2 are the only EEG segment lenghts that achieve the highest seizure detection accuracy of 100%. And since the EEG segment length of 2 yields a lower computational complexity than that of 1; we adopted this length in all our seizure detection experiments. In this regard, each EEG segment is designed to have only 2 data-points out of 4096, producing 2048 segments for each EEG channel signal.

*2) EEG Deep Feature Learning:*
In order to learn the expressive seizure characteristics from EEG data, deep learning was deployed to extract the discriminative EEG features pertinent to seizures. We design our deep neural network to include three layers, with a softmax classification layer on top of them. The EEG data samples were first passed through a fully connected LSTM layer of 100 neurons. The motivation for this was to learn the short and long term dependencies between the EEG segments in each signal and between the different EEG signals across the same class. Remembering information for long periods of time is practically the default behavior of LSTMs, making them the best candidate for processing long-term EEG signals.

As illustrated in Figure 2, the Dense layer was adopted to translate the information learned by the LSTM layer into meaningful seizure-associated features. And since our problem is a kind of sequence labeling problems, we deployed the time-distributed Dense layer (not the ordinary Dense layer) so that the cost function is calculated on all EEG time-steps and not the last one only. A fully-connected Dense layer of 50 units was used in this model.

The final structural step was to pass the output of the Dense layer through a 1D average pooling layer. The motivation for this was that all the EEG segments should contribute equally to the label prediction. The output of the Average Pooling layer is then presented as an input to the probabilistic classification model of softmax for EEG classification. The proposed deep learning model was trained and tested using two common scenarios: (1) The hold-out scenario: the EEG dataset was split into two sets, 80% of the data samples was used for training, and the remaining 20% was used for the classification[1]. (2) The cross-validation scenario: 3-folds, 5-folds, and 10-folds cross-validation were also used to train and test the proposed deep neural network.

*3) EEG Feature Classification:*
As shown in Figure 2, we add a softmax layer at the top of our model to generate label predictions. Softmax is the most common function used to represent a probability distribution in machine learning literature. From an optimization perspective, it has some subtle properties concerning differentiability. From a machine learning perspective: using a deep network with a softmax classifier on top can represent any K-class probability function over the feature space.

---

[1]Our experiments on the EEG feature learning using LSTM were conducted with the open-source software of Keras using TensorFlow backend [64].

In our EEG classification problem, the class labels are assumed to be: $y^{(i)} \in {1, \cdots, K}$, where $K$ is the total number of classes. Given a training set $\{(\mathbf{x}^{(1)}, y^{(1)}), (\mathbf{x}^{(2)}, y^{(2)}), \cdots, (\mathbf{x}^{(N)}, y^{(N)})\}$ of $N$ labeled samples, where $\mathbf{x}^{(i)} \in \Re^{(Q)}$. For each test sample $\mathbf{x}$, the softmax hypothesis evaluates the probability that $\mathcal{P}(y = k | \mathbf{x}(t), \mathbf{x}(t-1), \mathbf{x}(t-2), \cdots, \mathbf{x}(t-M))$ for each class label $k = 1, \cdots, K$; where $t$ represents the time-step shown in Figure 2 and $M$ is the total number of time-steps (segments). The summations of these $K$-probability values should equal to 1 and the highest probability belongs to the predicted class. Thus, the softmax hypothesis, denoted by $\boldsymbol{h}_\theta(\mathbf{x})$, is defined as follows:

$$\boldsymbol{h}_\theta(\mathbf{x}) = \begin{pmatrix} \mathcal{P}(y=1|\mathbf{x};\theta) \\ \mathcal{P}(y=2|\mathbf{x};\theta) \\ \vdots \\ \mathcal{P}(y=K|\mathbf{x};\theta) \end{pmatrix} = \frac{1}{\sum_{j=1}^{K} \exp(\theta_j^T \mathbf{x})} \begin{pmatrix} \exp(\theta_1^T \mathbf{x}) \\ \exp(\theta_2^T \mathbf{x}) \\ \vdots \\ \exp(\theta_K^T \mathbf{x}) \end{pmatrix}$$

where $\theta_1, \theta_2, \cdots, \theta_K$ are the softmax model parameters.

The cost function of the classifier is cross-entropy, denoted by $\boldsymbol{J}(\theta)$, described below:

$$\boldsymbol{J}(\theta) = -\left[\sum_{i=1}^{N}\sum_{k=1}^{K} \mathbb{1}\{y^{(i)} = k\} \log \mathcal{P}(y^{(i)} = k | \mathbf{x}^{(i)}; \theta)\right] \quad (11)$$

$$= -\left[\sum_{i=1}^{N}\sum_{k=1}^{K} \mathbb{1}\{y^{(i)} = k\} \log \frac{\exp(\theta_k^T \mathbf{x}^{(i)})}{\sum_{j=1}^{K} \exp(\theta_j^T \mathbf{x}^{(i)})}\right] \quad (12)$$

where $\mathbb{1}\{.\}$ is the "indicator function", which equals to 1 if the statement is true and 0 if the statement is false.

Then, an iterative optimization method such as the stochastic gradient descent [65], is used to minimize the cost function and maximize the probability of the correct class label.

The pseudo-code of the proposed LSTM-based seizure detection method is presented in Algorithm 1.

*4) Network Configuration:*
Our LSTM network was trained by optimizing the "categorical cross-entropy" cost function with "Adam" parameter update and a learning factor of $1 \times 10^{-3}$. The total number of LSTM units and Dense units was set to 100 and 50, respectively. The "return sequence" was set to "True" so that all EEG segments are considered in the feature extraction process. The batch sizes were set to 64 and the network parameters converged after around 2400 iterations with 40 epochs. The data were augmented by adding eye-blinking and muscle activity artifacts as well as Gaussian white noise, and various noise levels were considered in our experiments. Our implementation was derived in Python using Keras with TensorFlow backend and performed two hours training on a NVIDIA K40 GPU machine.

---

**Algorithm 1:** Epileptic Seizure Detection using Long-Short-Term Memory (ESD-LSTM).

1 **Input**: $Q$-dimensional EEG/iEEG Signal $\mathbf{x}$; Trained LSTM model
2 **Output**: Predicted EEG class label $\tilde{y} \to \{1, \cdots, K\}$
3 **Initialization**: $Q \leftarrow 4096$; $M \leftarrow 2048$;
4 **Initialization**: $K \leftarrow$ number of EEG classes; $K = 2, 3,$ and 5 for two-class, three-class, and five-class problems.
5 **procedure** ESD-LSTM($\mathbf{x}$, K, LSTM)
6 Pick an EEG segment length $L \in \{2^0, 2^1, 2^2, 2^3, \cdots, Q\}$;
7 Partitioning the EEG/iEEG signal into $M$ segments, each of $L$ length.
8 **while** $t \leq M$ **do**
9     $t \leftarrow t + 1$
10     $\mathbf{u}^t = \text{LSTM}(\mathbf{o}^t, \mathbf{c}^t, \mathbf{f}^t, \mathbf{i}^t, \mathbf{z}^t)$;     ▷ LSTM
11     $\mathbf{v}^t = \mathbf{h}_t(\mathbf{u}^t)$;     ▷ Dense
12 **end**
13 $\mathbf{E} = \mathbf{AP}(\mathbf{v}^t, \mathbf{v}^{t-1}, \mathbf{v}^{t-2}, \cdots, \mathbf{v}^{t-M})$;    ▷ Average Pooling
14 Compute $P_k = \{P_1, \cdots, P_K\} \leftarrow$ softmax($\mathbf{E}$)
15 Find Idx $\leftarrow$ Support(max($P_k$))    ▷ Index of highest probability
16 $\tilde{y} =$ Idx;    ▷ Predicted class label
17 **end procedure**

---

## V. RESULTS AND DISCUSSION

To evaluate the effectiveness of the proposed deep learning-based seizure detection approach, we compare its performance to those of the state-of-the-art detectors that use the same benchmark dataset. The detection performance was evaluated using the standard metrics, i.e., sensitivity (Sens), specificity (Spec), and classification accuracy (Acc).

### A. Seizure Detection in Ideal Conditions.

The proposed method is first examined in the ideal conditions, where the EEG recordings are assumed to be free of noise. The clean EEG signals are first segmented and then fed into the deep learning model with the specific goal of efficient EEG feature learning and classification.

*1) Two-class Classification Results:*
The first category of the two-class problems is to discriminate between the normal and seizure EEG epochs, which correspond to healthy and epileptic patients experiencing active seizures, respectively. The performance metrics of the proposed and relevant seizure detection methods are summarized in Table I. As shown in Table I, the sensitivity values are quite low for most of the seizure detectors reported in the literature. The highest sensitivity of 99.48% was achieved by Bugeja *et al.* using multilevel wavelet transform as a feature extraction method and extreme learning machine as a classification model [28]. It is interesting to find how clearly our seizure detection approach achieved a higher sensitivity of 100%.

Table I
SEIZURE DETECTION RESULTS OF THE PROPOSED AND STATE-OF-THE-ART METHODS: TWO-CLASS PROBLEM (A-E).

| Method | Year | Classifier | Training/Testing | Sens (%) | Spec (%) | Acc (%) |
|---|---|---|---|---|---|---|
| Aarabi et al. [18] | 2006 | BNN | Hold-out (50.00-50.00%) | 91.00 | 95.00 | 93.00 |
| Subasi et al. [19] | 2007 | ME | Hold-out (62.50-37.50%) | 95.00 | 94.00 | 94.50 |
| Chandaka et al. [21] | 2009 | SVM | Hold-out (62.50-37.50%) | 92.00 | 100.0 | 95.96 |
| Yuan et al. [22] | 2011 | ELM | Hold-out (50.00-50.00%) | 92.50 | 96.00 | 96.50 |
| Khan et al. [23] | 2012 | LDA | Hold-out (80.00-20.00%) | 83.60 | 100.0 | 91.80 |
| Nicolaou et al. [24] | 2012 | SVM | Hold-out (60.00-40.00%) | 94.38 | 93.23 | 93.80 |
| Zhou et al. [25] | 2013 | BLDA | Hold-out (95.00-05.00%) | 96.25 | 96.70 | 96.67 |
| Kumar et al. [26] | 2014 | SVM | Hold-out (33.33-66.67%) | 98.00 | 96.00 | 97.50 |
| Song et al. [27] | 2016 | SVM | – | 94.50 | 100.0 | 97.25 |
| Proposed Method | 2017 | Softmax | Hold-out (33.33-66.67%) | 100.0 | 100.0 | 100.0 |
| Bugeja et al. [28] | 2016 | ELM | Leave-one-out CV | 99.48 | 77.16 | – |
| Proposed Method | 2017 | Softmax | Leave-one-out CV | 100.0 | 100.0 | 100.0 |
| Polat et al. [20] | 2007 | DT | 10-folds cross-validation | 98.87 | 98.50 | 98.68 |
| Proposed Method | 2017 | Softmax | 10-folds cross-validation | 100.0 | 100.0 | 100.0 |

Table II
SEIZURE DETECTION RESULTS OF THE PROPOSED AND STATE-OF-THE-ART METHODS: TWO-CLASS PROBLEM (ABCD-E).

| Method | Year | Classifier | Training/Testing | Sens (%) | Spec (%) | Acc (%) |
|---|---|---|---|---|---|---|
| Guo et al. [29] | 2010 | ANN | Hold-out (50.00-50.00%) | 95.50 | 99.00 | 98.27 |
| Rivero et al. [30] | 2011 | KNN | Variable | – | – | 98.40 |
| Peker et al. [33] | 2016 | CVANN | Hold-out (60.00-40.00%) | 100.0 | 98.01 | 99.33 |
| Proposed Method | 2017 | Softmax | Hold-out (80.00-20.00%) | 100.0 | 100.0 | 100.0 |
| Kaleem et al. [31] | 2013 | KNN | 10-folds cross-validation | – | – | 98.20 |
| Fu et al. [32] | 2015 | SVM | 10-folds cross-validation | – | – | 98.80 |
| Jaiswal et al. [34] | 2017 | ANN | 10-folds cross-validation | 98.30 | 98.82 | 98.72 |
| Wang et al. [35] | 2017 | SVM | 10-folds cross-validation | 97.98 | 99.56 | 99.25 |
| Proposed Method | 2017 | Softmax | 10-folds cross-validation | 100.0 | 100.0 | 100.0 |

Further, our approach produced a notable seizure specificity of 100%, which is comparable to those of [23] and [27], and superior to those of the other baseline methods. Also, our approach can work on the raw EEG data and does not require any data pre-processing like those of [23] and [27]. Amongst all the existing seizure detection methods, the proposed approach yields superior classification accuracy of 100%, with a gap of 1.32% above the highest accuracy reported in the literature [20].

In the second evaluation, we address the classification problem between any non-seizure activities (sets A, B, C, or D) and seizure activities (set E). Given that each EEG set includes 100 signals, this classification problem has an unbalanced class distribution. This is because the number of EEG samples belonging to seizure class is significantly lower than those belonging to the non-seizure class. In this situation, the predictive model developed using conventional machine learning algorithms could be biased and inaccurate. Our approach, instead, can effectively address this kind of classification problems and beat the literature performance. Again, the performance is evaluated in terms of the sensitivity, specificity, and classification accuracy values. The performance metrics of the proposed and baseline methods are reported in Table II. They verify the superiority of the proposed approach over the state-of-the-art methods, while it achieves the topmost performance of 100% sensitivity, 100% specificity, and 100% classification accuracy.

*2) Three-class Classification Results:*

We also address the effectiveness of the proposed approach to distinguish between three different classes of EEG signals, which are *normal*, *inter-ictal*, and *ictal* EEGs. The classification performance of the proposed seizure detection method is compared to those of the state-of-the-art methods presented in [36]- [48]. All of these methods are examined on the same benchmark epileptic EEG dataset [15].

Table III comprises the performance metrics obtained by the proposed and the reference methods. It is clear that the proposed method outperforms all others in terms of sensitivity, specificity, and classification accuracy. The leading reason was using the LSTM that investigates the correlation between the EEG signals taken from different subjects and the dependencies between EEG segments of the same subject. The results shown in Table III demonstrate the high potential of deep neural networks to effectively learn the representative EEG features that best describe the behavior of normal, inter-ictal and ictal EEG activities. It is worth highlighting that the proposed approach yields a seizure sensitivity of 100%, which is superior to all the baseline methods. Further, the proposed method produces an eminent seizure specificity of 100%, which is similar to the recent results obtained by Acharya *et al.* [42], and is better than those of the reference methods. More interestingly, amongst other methods, the proposed approach achieves an outstanding classification accuracy of 100%.

*3) Five-class Classification Results:*

We also address the classification problem of the five different EEG sets of A, B, C, D, and E, respectively. This problem is more challenging than the above problems of 1) and 2) but has an advantage for many vital applications. It addresses the discrimination between EEG activities belonging to the same data class (e.g., sets C and D, which are both inter-ictai), aiming to provide more beneficial practices. For example, the classification between EEG sets C and D plays a key role in seizure localization, as their data were captured from different brain regions. Indeed, only few researchers paid attention to the importance of the five-class classification problem [49]-[54]. They, however, achieved adequate detection results, as shown in Table IV.

Table III
SEIZURE DETECTION RESULTS OF THE PROPOSED AND STATE-OF-THE-ART METHODS: THREE-CLASS PROBLEM (A-C-E).

| Method | Year | Classifier | Training/Testing | Sens (%) | Spec (%) | Acc (%) |
|---|---|---|---|---|---|---|
| Güler et al. [36] | 2005 | RNN | Hold-out (50.00-50.00%) | 95.50 | 97.38 | 96.79 |
| Tzallas et al. [37] | 2007 | ANN | Hold-out (50.00-50.00%) | 95.73 | 97.86 | 97.94 |
| Dastidar et al. [38] | 2008 | RBFNN | Hold-out (80.00-20.00%) | – | – | 96.60 |
| Übeyli et al. [39] | 2009 | MLPNN | Hold-out (50.00-50.00%) | 96.00 | 94.00 | 94.83 |
| Niknazar et al. [44] | 2013 | ECOC | Hold-out (70.00-30.00%) | 98.55 | 99.33 | 98.67 |
| Samiee et al. [46] | 2015 | MLPNN | Hold-out (50.00-50.00%) | 99.20 | 98.90 | 99.10 |
| Proposed Method | 2017 | Softmax | Hold-out (50.00-50.00%) | 100.0 | 100.0 | 100.0 |
| Hosseini et al. [47] | 2016 | SVM | Leave-one-out CV | 96.00 | 94.00 | 95.00 |
| Proposed Method | 2017 | Softmax | Leave-one-out CV | 100.0 | 100.0 | 100.0 |
| Nilchi et al. [41] | 2010 | MLPNN | 3-folds cross-validation | 97.46 | 98.74 | 97.50 |
| Acharya et al. [42] | 2012 | FSC | 3-folds cross-validation | 99.40 | 100.0 | 98.10 |
| Proposed Method | 2017 | Softmax | 3-folds cross-validation | 100.0 | 100.0 | 100.0 |
| Gajic et al. [45] | 2015 | PQ | 5-folds cross-validation | 98.60 | 99.33 | 98.70 |
| Proposed Method | 2017 | Softmax | 5-folds cross-validation | 100.0 | 100.0 | 100.0 |
| Song et al. [40] | 2010 | ELM | 10-folds cross-validation | 97.26 | 98.77 | 95.67 |
| Acharya et al. [43] | 2012 | GMM | 10-folds cross-validation | 99.00 | 99.00 | 99.00 |
| Behara et al. [48] | 2016 | LSSVM | 10-folds cross-validation | 96.96 | 99.66 | 97.19 |
| Proposed Method | 2017 | Softmax | 10-folds cross-validation | 100.0 | 100.0 | 100.0 |

Table IV
SEIZURE DETECTION RESULTS OF THE PROPOSED AND STATE-OF-THE-ART METHODS: FIVE-CLASS PROBLEM (A-B-C-D-E).

| Method | Year | Classifier | Training/Testing | Sens (%) | Spec (%) | Acc (%) |
|---|---|---|---|---|---|---|
| Güler et al. [49] | 2007 | PNN | Hold-out (50.00-50.00%) | 98.05 | 99.50 | 98.05 |
| Übeyli et al. [50] | 2008 | SVM | Hold-out (70.00-30.00%) | 99.30 | 99.82 | 99.30 |
| Übeyli et al. [51] | 2009 | SVM | Hold-out (50.00-50.00%) | 99.20 | 99.79 | 99.20 |
| Shen et al. [53] | 2013 | SVM | Hold-out (50.00-50.00%) | 98.37 | 100.0 | 99.97 |
| Siuly et al. [54] | 2014 | MSVM | Hold-out (50.00-50.00%) | 99.99 | 99.99 | 99.99 |
| Proposed Method | 2017 | Softmax | Hold-out (50.00-50.00%) | 100.0 | 100.0 | 100.0 |
| Murugavel et al. [52] | 2011 | MSVM | – | – | – | 96.00 |
| Proposed Method | 2017 | Softmax | 10-folds cross-validation | 100.0 | 100.0 | 100.0 |

We compare the performance of the proposed approach to the state-of-the-art methods that have been developed in the last decade. The performance metrics of all methods are reported in Table III. It is worth noting that the proposed method outperforms all others in terms of sensitivity, specificity, and classification accuracy. Comparing our results with the literature performance, we find that Siuly *et al.* developed a multiclass seizure detection method that achieves detection results comparable to those reported in our study, while it attains 99.99% sensitivity , 99.99% specificity, and 99.99% classification accuracy [54]. However, their method involves applying three pre-processing techniques, which are computationally intensive and might hinder the real-time applications. Our approach, on the other hand, relaxes the need of data pre-processing and works directly on the raw EEG data, achieving the superior detection performance of 100%.

*B. Seizure Detection in Real-life Conditions.*

We further examine the robustness of the proposed seizure detection method against the common EEG artifacts. In our previous work, we developed a reliable EEG feature learning method capable of performing on noisy signals [66]. This method, however, assumed that the only noise encountered during EEG acquisition has a Gaussian distribution, i.e., artifacts were excluded, which is not the case in practical situations. In this work, we introduce a practical seizure detection approach that can address noisy EEG data corrupted with real physical noise (muscle artifacts, eye-blinking and Gaussian white noise).

*1) Two-class Classification Results:*

We first investigate the performance of the proposed approach in recognizing whether the noise-corrupted EEG data correspond to a healthy person (set A) or an epileptic patient (set E). As shown in Figure 5, our method is examined at different noise levels. The common EEG artifacts of muscle activities and eye-blinking in addition to the white noise were considered, where their amplitudes were adjusted to produce noisy EEG signals of different SNRs. Figure 5 shows the seizure detection results obtained by our method in the presence of muscle activities, eye-blinking, and the white noise at a wide range of SNR ($-20$ to 20dB).

Several interesting observations can be made here. First, the proposed method can effectively learn the most discriminative and robust EEG features associated with seizures, even when the EEG data are completely immersed in noise. For example, Figure 5 demonstrates the robustness of our method in the presence of all sources of noise. Interestingly, for the noisy EEG corrupted by eye-blinking artifacts, the proposed method maintains a high classification accuracy of 100% at all SNR levels. The same applies to the noisy EEG contaminated with muscle artifacts and white noise, except when SNR=−20dB. The main reason was that, for SNR=−20dB, the EEG data were completely buried in noise and their original waveform shapes were distorted. The proposed method, however, preserves a high detection performance and achieves a classification accuracy of 99.75% and 99.25% for the case of muscle artifacts and white noise, respectively.

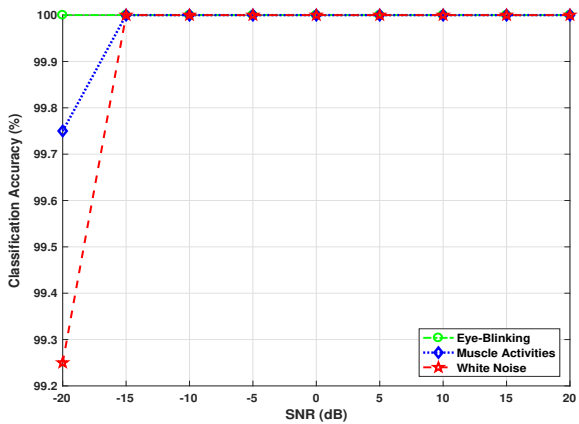

Figure 5. Classification accuracy vs. SNR plots for the two-class EEG classification problem (A-E).

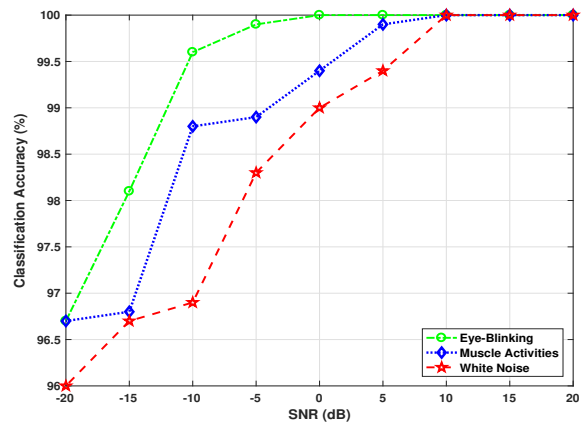

Figure 6. Classification accuracy vs. SNR plots for the two-class EEG classification problem (ABCD-E).

As for the two-class problem of ABCD-E, the proposed approach was also examined on noisy data contaminated with muscle artifacts, eye-blinking and electrical white noise. And since the dataset here is biased. i.e., it has an unbalanced class distribution, a negligible decay in the proposed method's performance was experienced. It is worth pointing out that, for such an unbalanced classification problem, the proposed method is proven to maintain high classification accuracies even at extremely low SNRs. Figure 6 illustrates the detection results obtained by our method in the presence of each noise type. It's clearly shown that the least classification accuracy of 96.70% was obtained when the EEG data was entirely immersed in white noise (SNR=−20). For noisy EEG data of SNR>0dB, the proposed method attains classification accuracies higher than 99.00%.

*2) Three-class Classification Results:*

Figure 7 investigates the performance of the proposed method in the presence of two common EEG artifacts and white noise at different SNR levels. It can be observed that the proposed method maintains its superior performance when applied to noise-corrupted data of SNRs above 0dB. The main reason is that LSTM networks can effectively learn the most discriminative and robust EEG features associated with seizures, even under noisy conditions. The performance of our model starts to decline when applied to noisy EEG data of SNRs below 0dB, particularly when the data is contaminated with white noise. A better performance can be achieved for the case of muscle artifacts since muscle activities interfere with EEG signals within a limited frequency band of 20-60Hz. A superior performance is achieved for the case of eye-blinking artifacts. Figure 7 verifies the robustness of the proposed approach against eye-blinking artifacts, even at extremely low SNRs. The proposed method can accurately identify seizure activities submerged in noise with acceptable classification accuracies.

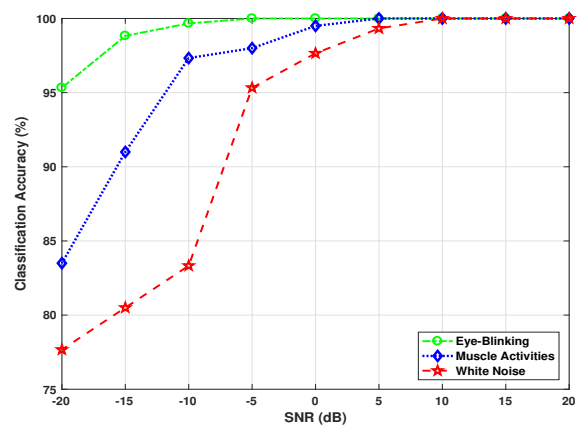

Figure 7. Classification accuracy vs. SNR plots for the three-class EEG classification problem (A-C-E).

*3) Five-class Classification Results:*

We also study the performance of the proposed seizure detection approach in the five-class classification problem under noisy conditions. This is when the EEG signals are mixed with different levels of muscle artifacts, eye-blinking, and white noise. Figure 8 demonstrates the detection performance of the proposed method at different SNRs. Even for this kind of intractable classification problem, the proposed approach is found to sustain classification accuracies higher than 94.00% for noisy EEG corrupted with eye-blinking artifacts. An inferior detection accuracy was obtained for the case of muscle artifacts; the classification accuracy is decreased to 70.90% at SNR=−20dB. The main reason is that muscle activities dwell in a wide range of EEG frequency spectrum producing a serious distortion in the EEG waveform shapes. Moreover, the performance of the proposed method encounters a high decay for the case of white noise; poor classification accuracies down to 53.50% were obtained. However, in more realistic situations (SNR>0dB), the proposed approach achieved superior performance with classification accuracies higher than 90.00%.

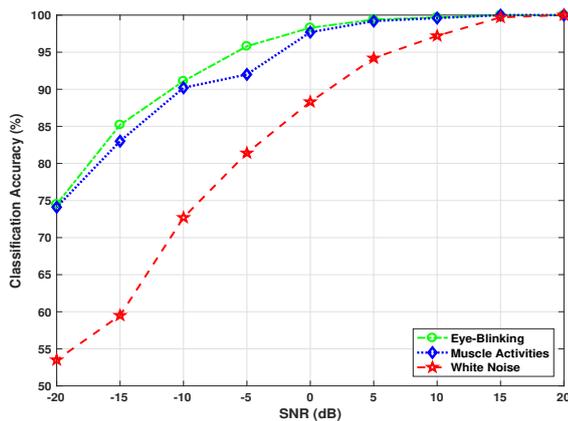

Figure 8. Classification accuracy vs. SNR plots for the five-class EEG classification problem (A-B-C-D-E).

## VI. CONCLUSION

In this paper, we introduce a deep learning approach for the automatic detection of epileptic seizures using EEG signals. Compared to the state-of-the-art methods, this approach can learn the high-level representations, and can effectively discriminate between the normal and seizure EEG activities. Another advantage of this approach lies in its robustness against common EEG artifacts (e.g., muscle activities and eye-blinking) and white noise. The proposed approach has been examined on the Bonn EEG dataset and compared to several baseline methods. The experimental results demonstrate the effectiveness and superiority of the proposed method in detecting epileptic seizures. It achieves the superior detection accuracies under ideal and imperfect conditions.

## REFERENCES

[1] G. Rogers, "Epilepsy: the facts," *Primary Health Care Research & Development*, vol. 11, no. 4, p. 413, 2010.
[2] U. R. Acharya, S. V. Sree, G. Swapna, R. J. Martis, and J. S. Suri, "Automated EEG analysis of epilepsy: a review," *Knowledge-Based Systems*, vol. 45, pp. 147–165, 2013.
[3] F. Mormann, R. G. Andrzejak, C. E. Elger, and K. Lehnertz, "Seizure prediction: the long and winding road," *Brain*, vol. 130, no. 2, pp. 314–333, 2006.
[4] R. Meier, H. Dittrich, A. Schulze-Bonhage, and A. Aertsen, "Detecting epileptic seizures in long-term human EEG: a new approach to automatic online and real-time detection and classification of polymorphic seizure patterns," *Journal of Clinical Neurophysiology*, vol. 25, no. 3, pp. 119–131, 2008.
[5] G. R. Minasyan, J. B. Chatten, M. J. Chatten, and R. N. Harner, "Patient-specific early seizure detection from scalp EEG," *Journal of clinical neurophysiology: official publication of the American Electroencephalographic Society*, vol. 27, no. 3, p. 163, 2010.
[6] A. G. Correa, E. Laciar, H. Patiño, and M. Valentinuzzi, "Artifact removal from EEG signals using adaptive filters in cascade," in *Journal of Physics: Conference Series*, vol. 90, no. 1. IOP Publishing, 2007, p. 012081.
[7] A. M. Chan, F. T. Sun, E. H. Boto, and B. M. Wingeier, "Automated seizure onset detection for accurate onset time determination in intracranial EEG," *Clinical Neurophysiology*, vol. 119, no. 12, pp. 2687–2696, 2008.
[8] A. Aarabi, R. Fazel-Rezai, and Y. Aghakhani, "A fuzzy rule-based system for epileptic seizure detection in intracranial EEG," *Clinical Neurophysiology*, vol. 120, no. 9, pp. 1648–1657, 2009.
[9] J. J. Niederhauser, R. Esteller, J. Echauz, G. Vachtsevanos, and B. Litt, "Detection of seizure precursors from depth EEG using a sign periodogram transform," *IEEE Transactions on Biomedical Engineering*, vol. 50, no. 4, pp. 449–458, 2003.
[10] I. Güler and E. D. Übeyli, "Adaptive neuro-fuzzy inference system for classification of EEG signals using wavelet coefficients," *Journal of neuroscience methods*, vol. 148, no. 2, pp. 113–121, 2005.
[11] A. T. Tzallas, M. G. Tsipouras, and D. I. Fotiadis, "Automatic seizure detection based on time-frequency analysis and artificial neural networks," *Computational Intelligence and Neuroscience*, vol. 2007, 2007.
[12] B. Abibullaev, H. D. Seo, and M. S. Kim, "Epileptic spike detection using continuous wavelet transforms and artificial neural networks," *International journal of wavelets, multiresolution and information processing*, vol. 8, no. 01, pp. 33–48, 2010.
[13] J. Mitra, J. R. Glover, P. Y. Ktonas, A. T. Kumar, A. Mukherjee, N. B. Karayiannis, J. D. Frost Jr, R. A. Hrachovy, and E. M. Mizrahi, "A multi-stage system for the automated detection of epileptic seizures in neonatal EEG," *Journal of clinical neurophysiology: official publication of the American Electroencephalographic Society*, vol. 26, no. 4, p. 218, 2009.
[14] K. Abualsaud, M. Mahmuddin, M. Saleh, and A. Mohamed, "Ensemble classifier for epileptic seizure detection for imperfect EEG data," *The Scientific World Journal*, vol. 2015, 2015.
[15] R. G. Andrzejak, K. Lehnertz, F. Mormann, C. Rieke, P. David, and C. E. Elger, "Indications of nonlinear deterministic and finite-dimensional structures in time series of brain electrical activity: Dependence on recording region and brain state," *Physical Review E*, vol. 64, no. 6, p. 061907, 2001.
[16] U. Herwig, P. Satrapi, and C. Schönfeldt-Lecuona, "Using the international 10-20 EEG system for positioning of transcranial magnetic stimulation," *Brain topography*, vol. 16, no. 2, pp. 95–99, 2003.
[17] A. Delorme, T. Sejnowski, and S. Makeig, "Enhanced detection of artifacts in EEG data using higher-order statistics and independent component analysis," *Neuroimage*, vol. 34, no. 4, pp. 1443–1449, 2007.
[18] A. Aarabi, F. Wallois, and R. Grebe, "Automated neonatal seizure detection: a multistage classification system through feature selection based on relevance and redundancy analysis," *Clinical Neurophysiology*, vol. 117, no. 2, pp. 328–340, 2006.
[19] A. Subasi, "EEG signal classification using wavelet feature extraction and a mixture of expert model," *Expert Systems with Applications*, vol. 32, no. 4, pp. 1084–1093, 2007.
[20] K. Polat and S. Güneş, "Classification of epileptiform EEG using a hybrid system based on decision tree classifier and fast fourier transform," *Applied Mathematics and Computation*, vol. 187, no. 2, pp. 1017–1026, 2007.
[21] S. Chandaka, A. Chatterjee, and S. Munshi, "Cross-correlation aided support vector machine classifier for classification of EEG signals," *Expert Systems with Applications*, vol. 36, no. 2, pp. 1329–1336, 2009.
[22] Q. Yuan, W. Zhou, S. Li, and D. Cai, "Epileptic EEG classification based on extreme learning machine and nonlinear features," *Epilepsy research*, vol. 96, no. 1-2, pp. 29–38, 2011.
[23] Y. U. Khan, N. Rafiuddin, and O. Farooq, "Automated seizure detection in scalp EEG using multiple wavelet scales," in *Signal Processing, Computing and Control (ISPCC), 2012 IEEE International Conference on*. IEEE, 2012, pp. 1–5.
[24] N. Nicolaou and J. Georgiou, "Detection of epileptic electroencephalogram based on permutation entropy and support vector machines," *Expert Systems with Applications*, vol. 39, no. 1, pp. 202–209, 2012.
[25] W. Zhou, Y. Liu, Q. Yuan, and X. Li, "Epileptic seizure detection using lacunarity and bayesian linear discriminant analysis in intracranial EEG," *IEEE Transactions on Biomedical Engineering*, vol. 60, no. 12, pp. 3375–3381, 2013.
[26] A. Kumar and M. H. Kolekar, "Machine learning approach for epileptic seizure detection using wavelet analysis of EEG signals," in *Medical Imaging, m-Health and Emerging Communication Systems (MedCom), 2014 International Conference on*. IEEE, 2014, pp. 412–416.
[27] Z. Song, J. Wang, L. Cai, B. Deng, and Y. Qin, "Epileptic seizure detection of electroencephalogram based on weighted-permutation entropy," in *Intelligent Control and Automation (WCICA), 2016 12th World Congress on*. IEEE, 2016, pp. 2819–2823.

[28] S. Bugeja, L. Garg, and E. E. Audu, "A novel method of EEG data acquisition, feature extraction and feature space creation for early detection of epileptic seizures," in *Engineering in Medicine and Biology Society (EMBC), 2016 IEEE 38th Annual International Conference of the*. IEEE, 2016, pp. 837–840.

[29] L. Guo, D. Rivero, and A. Pazos, "Epileptic seizure detection using multiwavelet transform based approximate entropy and artificial neural networks," *Journal of neuroscience methods*, vol. 193, no. 1, pp. 156–163, 2010.

[30] D. Rivero, E. Fernandez-Blanco, J. Dorado, and A. Pazos, "A new signal classification technique by means of genetic algorithms and knn," in *Evolutionary Computation (CEC), 2011 IEEE Congress on*. IEEE, 2011, pp. 581–586.

[31] M. Kaleem, A. Guergachi, and S. Krishnan, "EEG seizure detection and epilepsy diagnosis using a novel variation of empirical mode decomposition," in *Engineering in Medicine and Biology Society (EMBC), 2013 35th Annual International Conference of the IEEE*. IEEE, 2013, pp. 4314–4317.

[32] K. Fu, J. Qu, Y. Chai, and T. Zou, "Hilbert marginal spectrum analysis for automatic seizure detection in EEG signals," *Biomedical Signal Processing and Control*, vol. 18, pp. 179–185, 2015.

[33] M. Peker, B. Sen, and D. Delen, "A novel method for automated diagnosis of epilepsy using complex-valued classifiers," *IEEE journal of biomedical and health informatics*, vol. 20, no. 1, pp. 108–118, 2016.

[34] A. K. Jaiswal and H. Banka, "Local pattern transformation based feature extraction techniques for classification of epileptic EEG signals," *Biomedical Signal Processing and Control*, vol. 34, pp. 81–92, 2017.

[35] L. Wang, W. Xue, Y. Li, M. Luo, J. Huang, W. Cui, and C. Huang, "Automatic epileptic seizure detection in EEG signals using multi-domain feature extraction and nonlinear analysis," *Entropy*, vol. 19, no. 6, p. 222, 2017.

[36] N. F. Güler, E. D. Übeyli, and I. Güler, "Recurrent neural networks employing lyapunov exponents for EEG signals classification," *Expert systems with applications*, vol. 29, no. 3, pp. 506–514, 2005.

[37] A. T. Tzallas, M. G. Tsipouras, and D. I. Fotiadis, "Automatic seizure detection based on time-frequency analysis and artificial neural networks," *Computational Intelligence and Neuroscience*, vol. 2007, 2007.

[38] S. Ghosh-Dastidar, H. Adeli, and N. Dadmehr, "Principal component analysis-enhanced cosine radial basis function neural network for robust epilepsy and seizure detection," *IEEE Transactions on Biomedical Engineering*, vol. 55, no. 2, pp. 512–518, 2008.

[39] E. D. Übeyli, "Combined neural network model employing wavelet coefficients for EEG signals classification," *Digital Signal Processing*, vol. 19, no. 2, pp. 297–308, 2009.

[40] Y. Song and P. Liò, "A new approach for epileptic seizure detection: sample entropy based feature extraction and extreme learning machine," *Journal of Biomedical Science and Engineering*, vol. 3, no. 06, p. 556, 2010.

[41] A. R. Naghsh-Nilchi and M. Aghashahi, "Epilepsy seizure detection using eigen-system spectral estimation and multiple layer perceptron neural network," *Biomedical Signal Processing and Control*, vol. 5, no. 2, pp. 147–157, 2010.

[42] U. R. Acharya, F. Molinari, S. V. Sree, S. Chattopadhyay, K.-H. Ng, and J. S. Suri, "Automated diagnosis of epileptic EEG using entropies," *Biomedical Signal Processing and Control*, vol. 7, no. 4, pp. 401–408, 2012.

[43] U. R. Acharya, S. V. Sree, A. P. C. Alvin, and J. S. Suri, "Use of principal component analysis for automatic classification of epileptic EEG activities in wavelet framework," *Expert Systems with Applications*, vol. 39, no. 10, pp. 9072–9078, 2012.

[44] M. Niknazar, S. Mousavi, B. V. Vahdat, and M. Sayyah, "A new framework based on recurrence quantification analysis for epileptic seizure detection," *IEEE journal of biomedical and health informatics*, vol. 17, no. 3, pp. 572–578, 2013.

[45] D. Gajic, Z. Djurovic, J. Gligorijevic, S. Di Gennaro, and I. Savic-Gajic, "Detection of epileptiform activity in EEG signals based on time-frequency and non-linear analysis," *Frontiers in computational neuroscience*, vol. 9, p. 38, 2015.

[46] K. Samiee, P. Kovacs, and M. Gabbouj, "Epileptic seizure classification of EEG time-series using rational discrete short-time fourier transform," *IEEE transactions on Biomedical Engineering*, vol. 62, no. 2, pp. 541–552, 2015.

[47] M.-P. Hosseini, A. Hajisami, and D. Pompili, "Real-time epileptic seizure detection from EEG signals via random subspace ensemble learning," in *Autonomic Computing (ICAC), 2016 IEEE International Conference on*. IEEE, 2016, pp. 209–218.

[48] D. S. T. Behara, A. Kumar, P. Swami, B. K. Panigrahi, and T. K. Gandhi, "Detection of epileptic seizure patterns in EEG through fragmented feature extraction," in *Computing for Sustainable Global Development (INDIACom), 2016 3rd International Conference on*. IEEE, 2016, pp. 2539–2542.

[49] I. Guler and E. D. Ubeyli, "Multiclass support vector machines for EEG signals classification," *IEEE Transactions on Information Technology in Biomedicine*, vol. 11, no. 2, pp. 117–126, 2007.

[50] E. D. Übeyli, "Analysis of EEG signals by combining eigenvector methods and multiclass support vector machines," *Computers in Biology and Medicine*, vol. 38, no. 1, pp. 14–22, 2008.

[51] E. D. Übeyli, "Decision support systems for time-varying biomedical signals: EEG signals classification," *Expert Systems with Applications*, vol. 36, no. 2, pp. 2275–2284, 2009.

[52] A. M. Murugavel, S. Ramakrishnan, K. Balasamy, and T. Gopalakrishnan, "Lyapunov features based EEG signal classification by multi-class svm," in *Information and Communication Technologies (WICT), 2011 World Congress on*. IEEE, 2011, pp. 197–201.

[53] C.-P. Shen, C.-C. Chen, S.-L. Hsieh, W.-H. Chen, J.-M. Chen, C.-M. Chen, F. Lai, and M.-J. Chiu, "High-performance seizure detection system using a wavelet-approximate entropy-fsvm cascade with clinical validation," *Clinical EEG and neuroscience*, vol. 44, no. 4, pp. 247–256, 2013.

[54] Y. Li *et al.*, "A novel statistical algorithm for multiclass EEG signal classification," *Engineering Applications of Artificial Intelligence*, vol. 34, pp. 154–167, 2014.

[55] Y. Taigman, M. Yang, M. Ranzato, and L. Wolf, "Deepface: Closing the gap to human-level performance in face verification," in *Proceedings of the IEEE conference on computer vision and pattern recognition*, 2014, pp. 1701–1708.

[56] A. Krizhevsky, I. Sutskever, and G. E. Hinton, "Imagenet classification with deep convolutional neural networks," in *Advances in neural information processing systems*, 2012, pp. 1097–1105.

[57] H. Palangi, L. Deng, Y. Shen, J. Gao, X. He, J. Chen, X. Song, and R. Ward, "Deep sentence embedding using long short-term memory networks: Analysis and application to information retrieval," *IEEE/ACM Transactions on Audio, Speech and Language Processing (TASLP)*, vol. 24, no. 4, pp. 694–707, 2016.

[58] A. Graves, A.-r. Mohamed, and G. Hinton, "Speech recognition with deep recurrent neural networks," in *Acoustics, speech and signal processing (icassp), 2013 ieee international conference on*. IEEE, 2013, pp. 6645–6649.

[59] S. Hochreiter and J. Schmidhuber, "Long short-term memory," *Neural computation*, vol. 9, no. 8, pp. 1735–1780, 1997.

[60] J. Friedman, T. Hastie, and R. Tibshirani, "Regularization paths for generalized linear models via coordinate descent," *Journal of statistical software*, vol. 33, no. 1, p. 1, 2010.

[61] K. Greff, R. K. Srivastava, J. Koutník, B. R. Steunebrink, and J. Schmidhuber, "Lstm: A search space odyssey," *IEEE transactions on neural networks and learning systems*, vol. 28, no. 10, pp. 2222–2232, 2017.

[62] H. Azami, K. Mohammadi, and H. Hassanpour, "An improved signal segmentation method using genetic algorithm," *International Journal of Computer Applications*, vol. 29, no. 8, pp. 5–9, 2011.

[63] H. Hassanpour and M. Shahiri, "Adaptive segmentation using wavelet transform," in *Electrical Engineering, 2007. ICEE'07. International Conference on*. IEEE, 2007, pp. 1–5.

[64] R. Hussein, October 2017. [Online]. Available: https://github.com/ramyh/Epileptic-Seizure-Detection.git

[65] L. Bottou, "Large-scale machine learning with stochastic gradient descent," in *Proceedings of COMPSTAT'2010*. Springer, 2010, pp. 177–186.

[66] R. Hussein, Z. J. Wang, and R. Ward, "L1-regularization based EEG feature learning for detecting epileptic seizure," in *Signal and Information Processing (GlobalSIP), 2016 IEEE Global Conference on*. IEEE, 2016, pp. 1171–1175.